\documentclass[9pt,twocolumn,twoside]{pnas-new}

\usepackage{graphicx}
\usepackage{overpic}
\usepackage{bm}
\usepackage{tabularx}

\templatetype{pnasresearcharticle} 

\begin{document}

\title{Data-driven Mori-Zwanzig modeling of Lagrangian particle dynamics in turbulent flows}

\author[a,b,c,1]{Xander M. de Wit}
\author[b,c,d]{Alessandro Gabbana}
\author[b,c]{Michael Woodward}
\author[e]{Yen Ting Lin}
\author[a,f]{Federico Toschi}
\author[b,2]{Daniel Livescu}

\affil[a]{Fluids and Flows group and J.M. Burgers Center for Fluid Mechanics, Eindhoven University of Technology,
 5600 MB Eindhoven, Netherlands}
\affil[b]{Computational Physics and Methods Group (CCS-2), Los Alamos National Laboratory, Los Alamos, 87545 New Mexico,
 USA}
\affil[c]{Center for Nonlinear Studies (CNLS), Los Alamos National Laboratory, Los Alamos, 87545 New Mexico, USA}
\affil[d]{University of Ferrara and INFN Ferrara, 44121 Ferrara, Italy,
 USA}
\affil[e]{Information Science Group (CCS-3), Los Alamos National Laboratory, Los Alamos, 87545 New Mexico, USA}
\affil[f]{CNR-IAC, I-00185 Rome, Italy}

\leadauthor{de Wit}


\significancestatement{
Modeling the motion of small particles in turbulent flows is central to a wide range of scientific and engineering problems ranging from cloud formation and pollutant transport to manufacturing and nuclear fusion. Yet, despite a century of research, reproducing the statistical features of these Lagrangian trajectories has remained a formidable challenge for traditional modeling approaches. We introduce a data-driven auto-regressive framework that captures the dynamics of Lagrangian turbulence within a surrogate dynamical system. This advance demonstrates that machine learning can overcome longstanding barriers in modeling chaotic particle motion, opening new opportunities for prediction, design, and control of turbulent processes across science and engineering.}

\authordeclaration{The authors declare no competing interests.}
\equalauthors{\textsuperscript{1}E-mail: x.m.d.wit@tue.nl}
\correspondingauthor{\textsuperscript{2}E-mail: livescu@lanl.gov}


\begin{abstract}
The dynamics of Lagrangian particles in turbulence play a crucial role in mixing, transport, and dispersion in complex flows. Their trajectories exhibit highly non-trivial statistical behavior, motivating the development of surrogate models that can reproduce these trajectories without incurring the high computational cost of direct numerical simulations of the full Eulerian field. This task is particularly challenging because reduced-order models typically lack access to the full set of interactions with the underlying turbulent field. Novel data-driven machine learning techniques can be powerful in capturing and reproducing complex statistics of the reduced-order/surrogate dynamics. 
In this work, we show how one can learn a surrogate dynamical system that is able to evolve a turbulent Lagrangian trajectory in a way that is point-wise accurate for short-time predictions (with respect to Kolmogorov time) and stable and statistically accurate at long times. This approach is based on the Mori-Zwanzig formalism, which prescribes a mathematical decomposition of the full dynamical system into resolved dynamics that depend on the current state and the past history of a reduced set of observables, and the unresolved orthogonal dynamics due to unresolved degrees of freedom of the initial state. We show how by training this reduced order model on a point-wise error metric on short time-prediction, we are able to correctly learn the dynamics of Lagrangian turbulence, such that also the long-time statistical behavior is stably recovered at test time. This opens up a range of new applications, for example, for the control of active Lagrangian agents in turbulence.
\end{abstract}

\dates{This manuscript was compiled on \today}
\doi{\url{www.pnas.org/cgi/doi/10.1073/pnas.XXXXXXXXXX}}

\maketitle
\thispagestyle{firststyle}
\ifthenelse{\boolean{shortarticle}}{\ifthenelse{\boolean{singlecolumn}}{\abscontentformatted}{\abscontent}}{}

\firstpage[6]{4}

\dropcap{U}nderstanding and predicting the trajectories of particle tracers advected by fully developed turbulent flows is a central challenge in fluid dynamics, with applications spanning a wide range of fields, including pollutant dispersion, cloud microphysics, aerosol dynamics in atmospheric chemistry, industrial processes, combustion processes, and biophysical systems~\cite{pope-book-2001, yeung-arfm-2002, falkovich-nat-2002}. 
Progress in this direction has proven difficult due to the chaotic and multiscale nature of turbulence, as the dynamics of particles is influenced by complex interactions across a wide range of spatial and temporal scales~\cite{sawford-arfm-2001,bec-pof-2006,toschi-arfm-2009}.

A natural framework for studying this problem is the Lagrangian description of fluid motion, where the focus is on following individual particles as they are transported by the fluid flow. In this context, Lagrangian tracers represent idealized point-like and massless particles, whose instantaneous velocities match the local Eulerian velocity field $\bm{u}(\bm{r}, t)$, such that their dynamics obeys $ \bm{v}(t) \equiv \dot{\bm{x}}(t) = \bm{u}(\bm{x}(t), t)$. 
As Lagrangian tracers travel the entire range of multiscale turbulent fluctuations, they recover many of the hallmarks of the statistics of turbulence~\cite{toschi-arfm-2009}, as the Lagrangian and Eulerian descriptions are closely connected. In particular, the small-scale dynamics of velocity structures is associated with particle accelerations, which are highly intermittent and exhibit extreme fluctuations. This manifests itself most strikingly through the probability density functions (PDFs) of Lagrangian accelerations encompassing heavy, non-Gaussian tails~\cite{laporta-nat-2001,voth-jfm-2002,mordant-physD-2004}.

From the numerical point of view, turbulent Lagrangian dynamics is typically studied using direct numerical simulations (DNS).
There, Lagrangian trajectories of particles in turbulent flows are computed from the fully resolved velocity field, which involves solving the full Navier--Stokes equations numerically without modeling approximations~\cite{yeung-jcp-1988,yeung-arfm-2002,toschi-arfm-2009}. 

While DNS provides the most accurate trajectories, the associated high computational costs have motivated the development of a wide range of reduced order models over the past decades, which in general do not require the explicit solution of the underlying flow field, and yet are successful in preserving (some of) the statistical features of Lagrangian turbulence.
Many of such approaches rely on stochastic models in which, following a Langevin-like approach, particle velocity and/or acceleration are modeled as a stochastic process, constrained \textit{a priori} to comply with certain turbulence statistics~\cite{sawford-pofa-1991}.
Extensions based on multifractal models further refine these approaches by embedding intermittency effects~\cite{wilson-blm-1996, biferale-pof-2005, lamorgese-jfm-2007, minier-pof-2014,viggiano-jfm-2020} or multi-particle correlations~\cite{mazzitelli2014accurate}.
Other notable alternatives are given by kinematic simulations (KS)~\cite{fung-jfm-1992,murray-pof-2016}, which generate synthetic velocity fields as a superposition of random Fourier modes with prescribed energy spectra, and shell models, which provide a reduced-order representation of turbulent energy transfer in Fourier space, which can in turn be translated to the Lagrangian viewpoint~\cite{biferale-arfm-2003}.

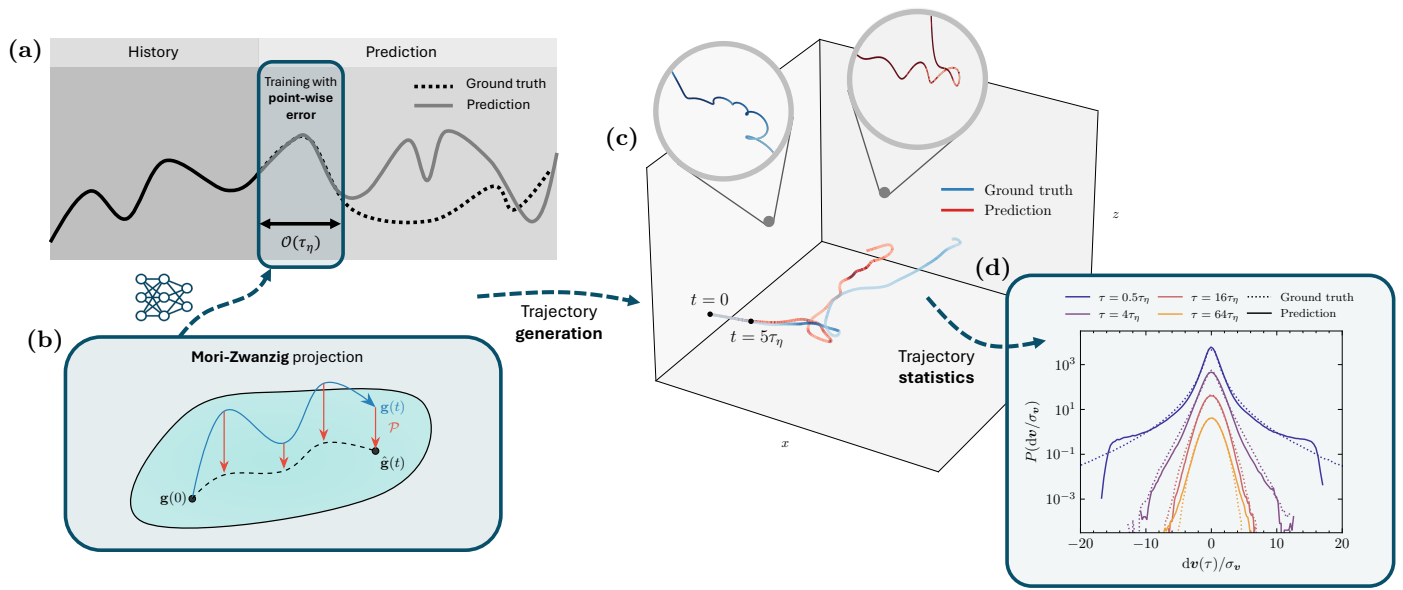
\begin{figure*}
   \centering
   \begin{overpic}[width=\linewidth]{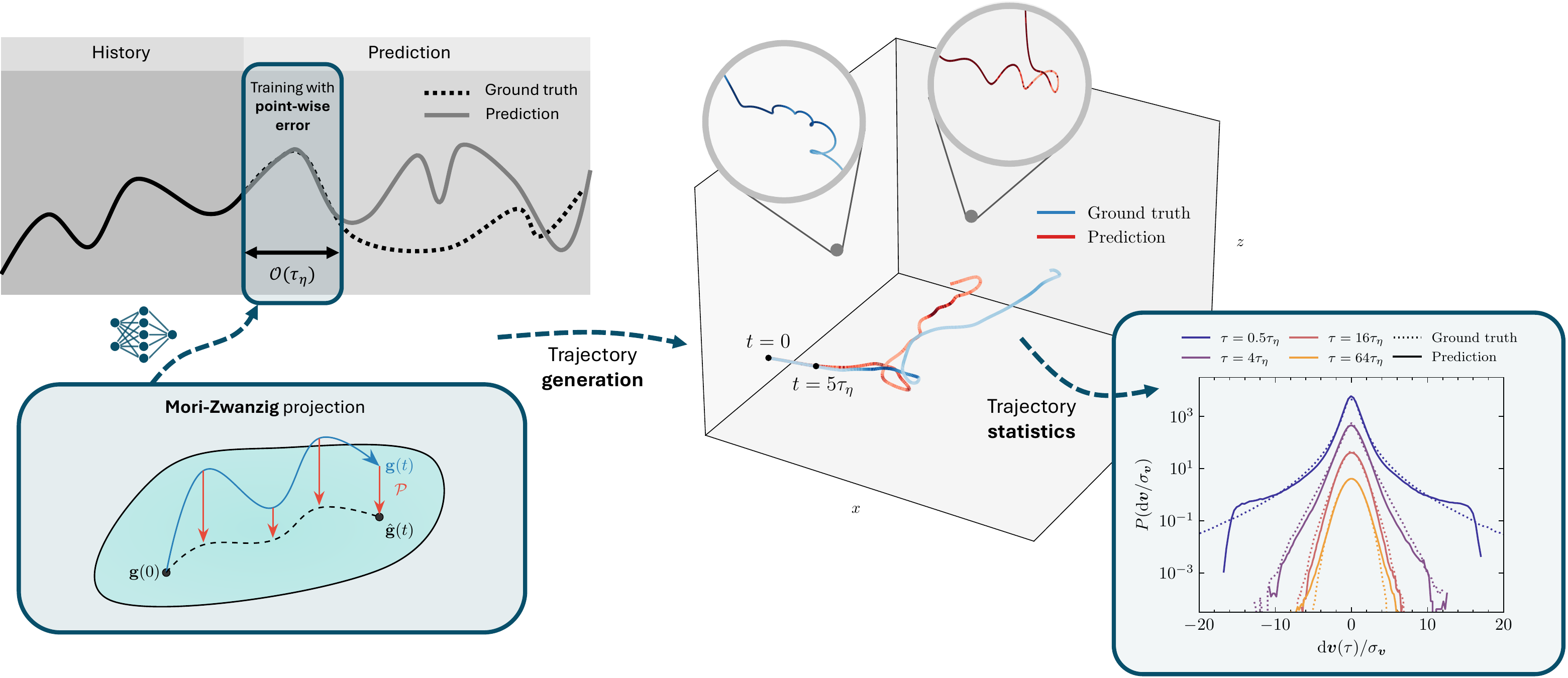}
     \put(-3.8,  39.5){\textbf{ (a) }}
     \put(-2.4,  18){\textbf{ (b) }}
     \put(40.5,  33){\textbf{ (c) }}
     \put(67.9,  23.3){\textbf{ (d) }}
   \end{overpic}
   \caption{ (a) Schematic depiction of the learning task for the Mori-Zwanzig (MZ) surrogate model: feeding a limited
      ground-truth history, the MZ model aims to predict the continuation of the trajectory in a way that is
      point-wise accurate at short times (w.r.t. Kolmogorov time $\tau_\eta$) and statistically accurate at longer
      times. (b) Sketch of the data-driven MZ approach: we begin with the true evolution of the observables $\bm g(t)$ described by the generalized Langevin equation (GLE, \eqref{eq:GLE}). Applying the projection operator $\mathcal{P}$ eliminates the orthogonal‑dynamics contribution, so the resulting GLE is formulated solely in the projected subspace $\hat{\bm{g}} := \mathcal{P}\bm g$. (c) Exemplary 3D Lagrangian trajectories, comparing ground-truth data (blue)
      with its corresponding MZ prediction (red). Trajectories are colored by the local vorticity magnitude (darker color means larger vorticity). One can appreciate the development of intermittent vortex-filament structures in certain trajectories (insets). (d) PDFs of Lagrangian velocity differences, normalized by their standard deviation, for different time lags $\tau$. PDFs are vertically shifted for readability.
      }\label{fig:schematic}
\end{figure*}

Although these modeling approaches can capture certain statistical features of turbulence, they often fail to reproduce the full complexity of Lagrangian turbulent statistics as observed in DNS. This has motivated the development of reduced-order models based on data-driven approaches~\cite{friedrich-jpc-2021,sinhuber-njp-2021} in recent years, enabled by the increasing availability of high-resolution data from DNS~\cite{kaneda-pof-2003,yeung-pnas-2015,yeung-prf-2020} and experimental measurements, such as those obtained through interpolated Particle Image Velocimetry (PIV)~\cite{schanz-ef-2016,machicoane-fm-2019,schroeder-arfm-2023}. 
Notable examples include diffusion models, which have shown remarkable success in generating synthetic trajectories while preserving key statistical properties, including accurate multi-scale properties beyond the restricted range where pure power laws are observed~\cite{li-nmi-2024,li-ijmf-2024}. Such data-driven approaches thereby outperform traditional approaches based on stochastic models in terms of fidelity of the statistics.
However, existing diffusion-based approaches still face important limitations, particularly because of the absence of an explicit underlying dynamical evolution. Specifically, diffusion models -- being probabilistic generative models -- synthesize entire trajectories through a stochastic sampling procedure, rather than learning the time-evolution equations that govern the underlying stochastic process. As a result, while these models are effective for trajectory synthesis, their applicability to problems requiring a true (surrogate) dynamical system that can be used in an auto-regressive way remains an open challenge.

In this work, we utilize a mathematical formalism based on non-equilibrium statistical mechanics and dynamical systems theory for the data-driven discovery of an auto-regressive reduced Lagrangian model for turbulence. 
The method introduced here is capable of learning a surrogate dynamical system which, provided with a short time history of the system, allows for accurate point-wise predictions within time scales comparable to the Kolmogorov scale $\tau_{\eta}$, and remains stable and statistically accurate on longer time scales (cf.~Fig.~\ref{fig:schematic}a).
The backbone of the method relies on a unified data-driven approach combining the Mori--Zwanzig formalism~\cite{mori1965transport, zwanzig1973nonlinear} and Takens' time-delay embedding~\cite{takens81}, two established memory embeddings for the systematic development of closure modeling.

Crucially, we show that training the model over relatively short time intervals, on the order of the Kolmogorov time scale, and using a simple point-wise loss function such as the mean squared error, suffices to capture the underlying physical mechanisms driving the dynamics.
Then, because the dynamics is correctly learned, at test time, the model is able to accurately recover the long-time statistics of turbulent Lagrangian accelerations and velocity gradients, including their probability distributions and \mbox{(auto-)correlation} structures, without incorporating any explicit statistical constraints in the training process.

We further highlight the robustness of the approach in addressing a common limitation of auto-regressive, data-driven models: long-term stability. Despite being trained only on short trajectories, the model produces statistically accurate and stable predictions over significantly longer time horizons, even when initialized with highly out-of-distribution unphysical states (e.g., zero velocity perturbed by random noise).

\subsection*{Training dataset}\label{sec:dataset}

We train the model on a comprehensive dataset of Lagrangian trajectories recently introduced in Ref.~\cite{biferale-arxiv-2023}. The data were obtained from direct numerical simulations of the three-dimensional incompressible Navier-Stokes equations under conditions of statistically stationary, homogeneous, and isotropic turbulence. The simulations were performed in a triply periodic cubic domain of size $L = 2\pi$, discretized on a $1024^3$ grid employing a standard pseudo-spectral method with full dealiasing~\cite{de2024efficient}. The database provides high-resolution time-resolved measurements of particle positions, velocities, and accelerations, along with the velocity gradient tensor evaluated at the particle locations. Data are sampled over a total time span of $T = 195\,\tau_{\eta}$, with a temporal resolution of $\Delta t = 0.1\,\tau_\eta$, where $\tau_\eta$ denotes the Kolmogorov timescale, which is the fastest dynamical timescale in turbulence. Our proposed model used the same timestep in order to resolve all dynamically active timescales. The corresponding Taylor-scale Reynolds number is $\rm{Re}_\lambda \approx 310$, with estimated Lagrangian integral timescale $T_L \approx 310\,\tau_\eta$.

The full dataset is split into disjoint sets: a training set (65\%) used for optimization of the model weights, a validation set (28\%) used for early stopping to avoid overfitting at training time and a test set (7\%) that is exclusively used for evaluation of the model after training, as laid out in the section \textit{Model evaluation}. Trajectories in the training set and validation set are furthermore split in time disjointly, since only short roll-outs need to be performed at training time. This removes any potential explicit temporal bias in the training set.

\subsection*{Model architecture and training}\label{sec:architecture}

The model is based on a dynamical equation with a finite memory that is used to account for the unresolved dynamics. It evolves the particle acceleration $\bm{a}(t)$ and local velocity gradients $\bm{\nabla}\bm{u}(t)$ at the position of the particle, while the particle velocity $\bm{v}(t)$ and position $\bm{x}(t)$ are recovered through integration. The auto-regressive kernels are parameterized with fully connected neural networks that are regressed on the training data.

Our method follows the principled approaches to extract Markov and memory kernels in the Mori-Zwanzig formalism, established in Refs.~\cite{lin-siam-2023,woodward23_aviation}. In short, training the Mori-Zwanzig model with time-delay embeddings is done in a cascaded fashion. The first MZ kernel (i.e. the Markov term) is fitted by minimizing the error after taking in the current state and its time-delay embeddings and predicting the next time step. Then, keeping the first MZ kernel fixed, the residual error is minimized by fitting the second MZ kernel, which takes in the state that is time-shifted into the history by one step. This process repeats for the full MZ memory length. Essential to the MZ formalism is that this recursive procedure is based on the satisfaction of the Dissipation-Fluctuation Theorem which relates the two-time correlation of the residual to the memory kernels.

The complete details on the derivation of the model and its architecture, the choice of hyper-parameters and the training procedure are provided in the \textit{Methods}.

\section*{Results}\label{sec:numerics}

\subsection*{Model evaluation}\label{sec:evaluation}

\begin{figure}
    \centering
    \includegraphics[width=.9\linewidth]{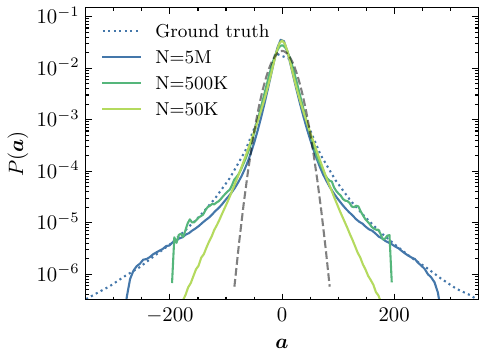}
    \caption{Convergence of the accuracy of the predicted PDF of particle acceleration with increasing number of trajectories in the training set from 50K to 5M. Dashed line indicates a Gaussian distribution with the same variance.
    }
    \label{fig:convergence}
\end{figure}

To evaluate the performance of the surrogate model, we begin by assessing the convergence of the model predictive accuracy as a function of the size of the training set.  Fig.~\ref{fig:convergence} shows the PDF of particle acceleration, comparing the results obtained when training over 50K, 500K and 5M trajectories. As evident from the figure, increasing the amount of training data leads to systematic improvements in the quality of the predictions, particularly in the accurate reconstruction of the heavy tails of the PDF, which are associated with extreme acceleration events. Based on this analysis, all subsequent results are reported for the model trained on the largest dataset of 5M trajectories.

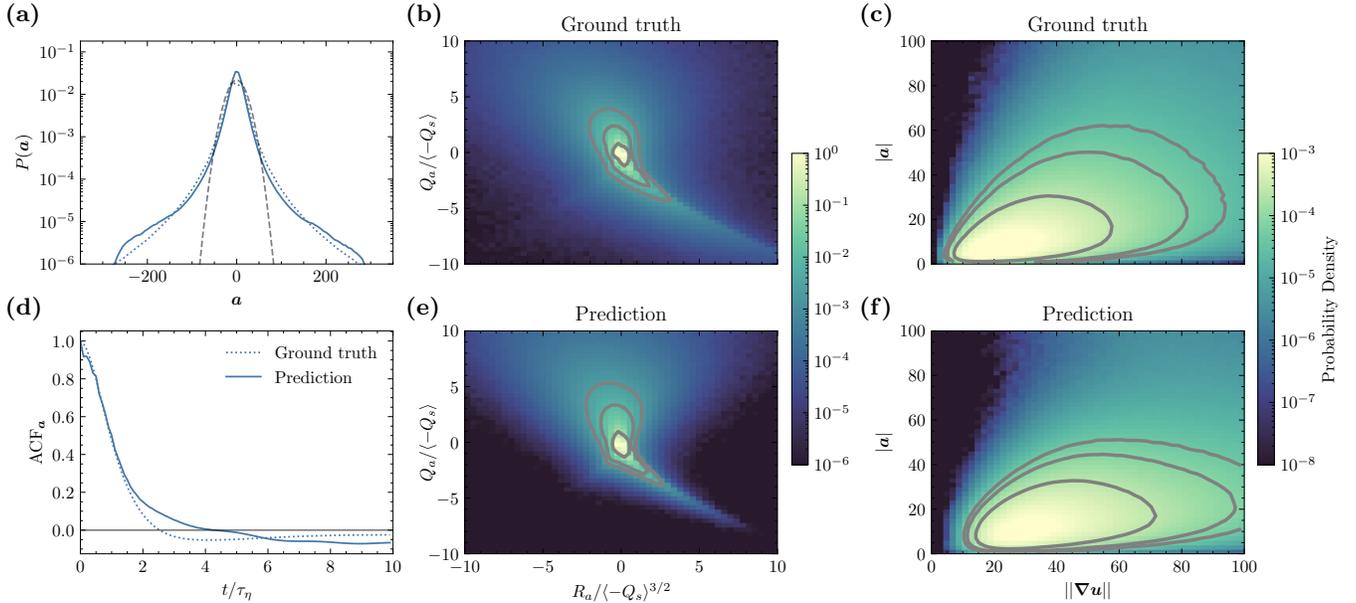
\begin{figure*}[htb]
    \centering
    \setlength{\fboxrule}{1pt}
    \begin{overpic}[width=.99\linewidth]{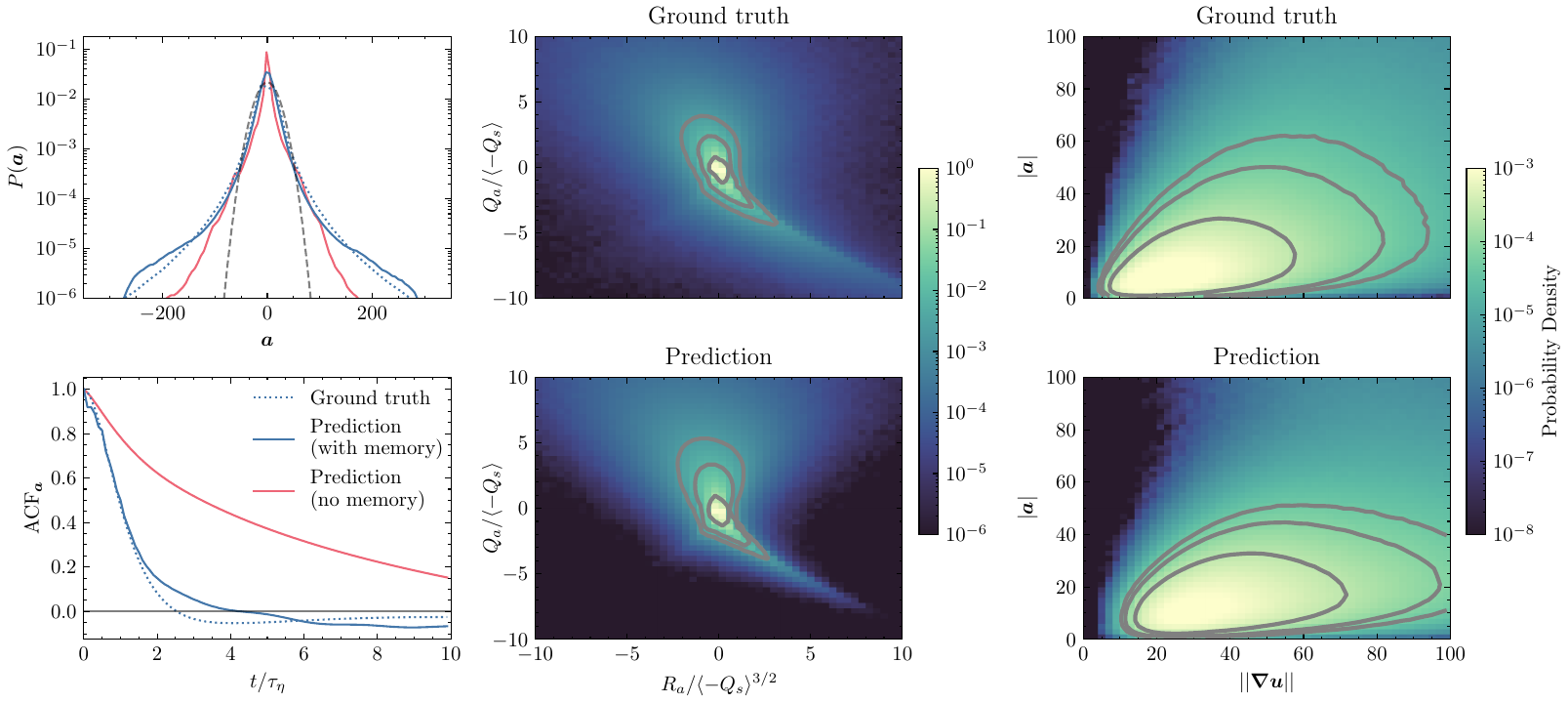}
     \put(-1,  44){\textbf{ (a) }}
     \put(29,  44){\textbf{ (c) }}
     \put(63,  44){\textbf{ (e) }}
     \put(-1,   22){\textbf{ (b) }}
     \put(29,   22){\textbf{ (d) }}
     \put(63,   22){\textbf{ (f) }}
    \end{overpic}
    \caption{Summary statistics of the long time prediction of the MZ model, compared with the ground truth. Provided are the acceleration PDF (a), its autocorrelation (b), the Q-R diagram of the velocity gradient at the position of the particle (c,d) and the joint PDF of the magnitude of acceleration and velocity gradient (e,f). The dashed line in (a) indicates a Gaussian distribution with the same variance. The lines in (c-f) correspond to level contours of 70\%, 90\% and 95\% probability percentiles.
    }\label{fig:best_stats}
\end{figure*}

We then evaluate the statistical performance of the model on key observables in the context of the dynamics of Lagrangian particles in turbulent flows. As highlighted in the introduction, the PDF of particle accelerations exhibits some of the most extreme fluctuations found in turbulence. In Fig.~\ref{fig:best_stats}a we compare the reference PDF computed over the entire dataset of Ref.~\cite{biferale-arxiv-2023} with the long-term predictions on $\mathcal{O}(10^2 \tau_{\eta})$ generated by our model. The model faithfully reproduces the heavy tails of the distribution, indicating its ability to capture the intermittent and strongly non-linear dynamics underlying particle acceleration. We find that the 2nd moment agrees within 2\% and the 4th moment agrees within 14\%.

In Fig.~\ref{fig:best_stats}d we take into consideration the particle acceleration autocorrelation function, formally defined as 
\begin{equation}
    \textrm{ACF}_{\bm{a}} = \frac{\left\langle \bm{a}(t_0+t) \cdot \bm{a}(t_0)\right\rangle}{\left\langle |\bm{a}(t_0)|^2 \right\rangle}
\end{equation}
with $\langle ...\rangle$ denoting an ensemble average over all particles and over time origin $t_0$.
The results are in satisfactory agreement with the ground-truth, indicating that the statistics of the temporal structure of the Lagrangian acceleration are properly captured by the surrogate model. A similar analysis can be performed for the predicted velocity gradient, and is reported in the SI Appendix~C.

To investigate the importance of memory, we compare the prediction of our model with Mori-Zwanzig memory and time-delay embedding to a model that has no memory in Fig~\ref{fig:best_stats}a,d. To ensure a fair comparison, the memoryless model has the same number of weights, the same architecture and the same number of kernels, except that all kernels are purely Markovian and the state is not enhanced with time-delay embeddings. The results show that such a memoryless model falls short in terms of fidelity of the resulting distribution and particularly cannot capture the correct temporal correlation structure of the dynamics. In SI Appendix~D, we show that even when further increasing the network size of the memoryless model, it cannot get appreciably closer to the performance of the memory-based approach. This underpins the principal importance of including memory in order to capture the effect of the unresolved dynamics in the reduced order model.

In order to evaluate the quality of the predictions for the fluid velocity gradient, in Fig.~\ref{fig:best_stats}b,e we show the so-called Q-R plot, a diagnostic tool commonly used to study the local topology of turbulence, in particular the relationship between the rate of strain and vorticity.
For this analysis, we start by introducing a decomposition of the velocity gradient tensor $\nabla \bm{u} = \bm{S} + \bm{A}$ into its symmetric part $\bm{S}$, the strain-rate tensor, and its antisymmetric part $\bm{A}$, the rotation-rate tensor, with
\begin{equation}
   \bm{S} = \tfrac{1}{2}(\nabla \bm{u} + (\nabla \bm{u})^T), 
   \quad 
   \bm{A} = \tfrac{1}{2}(\nabla \bm{u} - (\nabla \bm{u})^T).
\end{equation}
Based on this decomposition, we introduce the quantities
\begin{equation}
   Q_s = -\tfrac{1}{2} \mathrm{Tr}[\bm{S}^2], 
   \quad 
   Q_a = -\tfrac{1}{2} \mathrm{Tr}[\bm{A}^2], 
   \quad 
   R_a = -\tfrac{1}{3} \mathrm{Tr}[\bm{A}^3].
\end{equation}
Here, $Q_s$ represents the local rate of strain intensity, $Q_a$ measures the local rotation intensity, and $R_a$ characterizes the third-order moment of the rotational component.
In the figure, the quantities $Q_a$ and $R_a$ are normalized by suitable powers of the average strain contribution $\langle -Q_s \rangle $ (since $Q_s$ is negative definite in incompressible HIT), averaging over different time stamps and over particles, such to define non-dimensional quantities, and plotted as joint PDF.
We observe a good qualitative agreement with ground-truth data, with both figures highlighting the characteristic teardrop shape, reflecting the competition between vortical and straining motions. This offers a convincing validation for the physical fidelity of the predicted velocity gradients.
Additionally, we verify whether the correlation between acceleration and velocity gradient is properly captured. To that extent, in Fig.~\ref{fig:best_stats}c,f we analyze joint PDF of the magnitude of acceleration $|\bm{a}|$ and of the velocity gradient $||\bm{\nabla}\bm{u}||$. The figure shows good qualitative agreement between the joint PDF in the prediction and the ground truth. Indeed, events of high velocity gradient are positively correlated with high acceleration, indicating that when the particle encounters e.g. a region of high vorticity, it will experience a large acceleration and vice-versa, which is well resolved by the surrogate model.

The accurate reproduction of the heavy-tailed acceleration distribution, the proper time-correlation structure, and the correct velocity gradient statistics provide convincing evidence that the surrogate model is capable of faithfully capturing the complex statistical properties of Lagrangian turbulence, and most notably, its intermittency. As a consequence, at the more qualitative level, in examples of the generated trajectories, one can observe instances that resemble encounters with vortex filaments -- a feature often associated with turbulence intermittency (see example in Fig.~\ref{fig:schematic}c).

\begin{figure}[tb]
    \centering
    \setlength{\fboxrule}{1pt}
    \begin{overpic}[width=.95\linewidth]{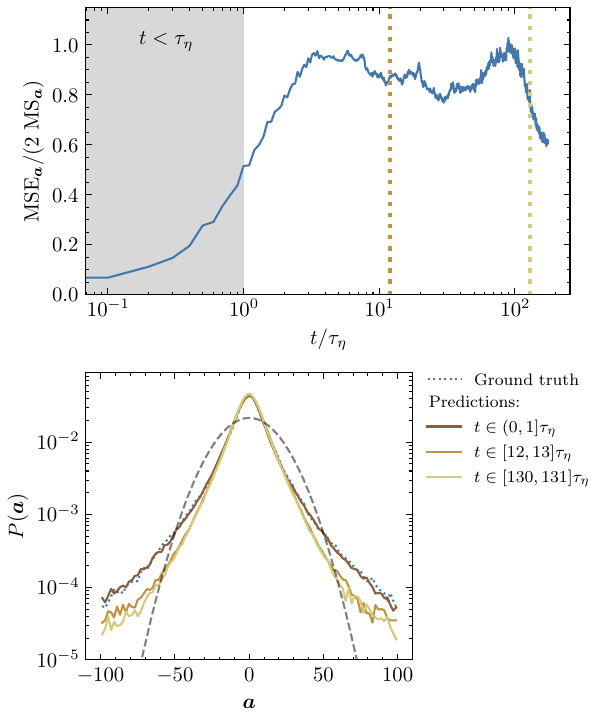}
     \put(-1,  100){\textbf{ (a) }}
     \put(-1,  50){\textbf{ (b) }}
    \end{overpic}
    \vspace{-1mm}
    \caption{Temporal robustness of the MZ model. (a) the mean-squared error (MSE) of acceleration, normalized by twice the mean square (MS) acceleration, as a function of time. (b) PDFs of acceleration computed over different time windows. The dashed line indicates a Gaussian distribution with the same variance.}
    \label{fig:time_dependence}
\end{figure}

In Fig.~\ref{fig:time_dependence}, we investigate the temporal robustness of the model predictions. The upper panel shows the MSE computed at fixed time steps, with time on the horizontal axis scaled by $\tau_{\eta}$. Predictions are compared against the ground truth, where the model is supplied with memory terms extracted from the corresponding ground truth trajectories. Initially, the MSE grows gradually up to approximately one Kolmogorov time, consistent with the design of the loss function used for training (\eqref{eq:nn-loss}). Beyond this range, the error increases before eventually saturating and fluctuating around a steady value at large times. Indeed, owing to the chaotic nature of turbulence, it is not feasible to achieve point-wise agreement between predicted and true trajectories much beyond the Kolmogorov time scale.

In the lower panel, we present the PDF of the acceleration evaluated over three selected time windows. As anticipated from the MSE behavior, the PDF computed over the interval $(0,1)\tau_{\eta}$ shows the best agreement with the ground truth, as all trajectories start from the ground truth statistics that result from the true turbulent attractor. Crucially, the figure shows that statistically, although the model was trained only on short trajectories within $(0,0.5)\tau_{\eta}$, it continues to yield accurate statistics even at moderate, $(12,13)\tau_{\eta}$, and long, $(130,131)\tau_{\eta}$, time scales. This result indicates that the model successfully captures the essential physical mechanisms governing the system dynamics, despite being trained on a short time horizon. It shows that as the dynamics of the system evolves, it is able to keep the trajectories close to the true turbulent attractor, thereby producing the correct Lagrangian statistics. This confirms that the model achieves high accuracy on the point-wise prediction over short time scales, of the order of the Kolmogorov time $\tau_{\eta}$, while sustaining statistically reliable behavior at longer times.

\begin{figure*}
    \centering
    \begin{overpic}[width=.99\linewidth]{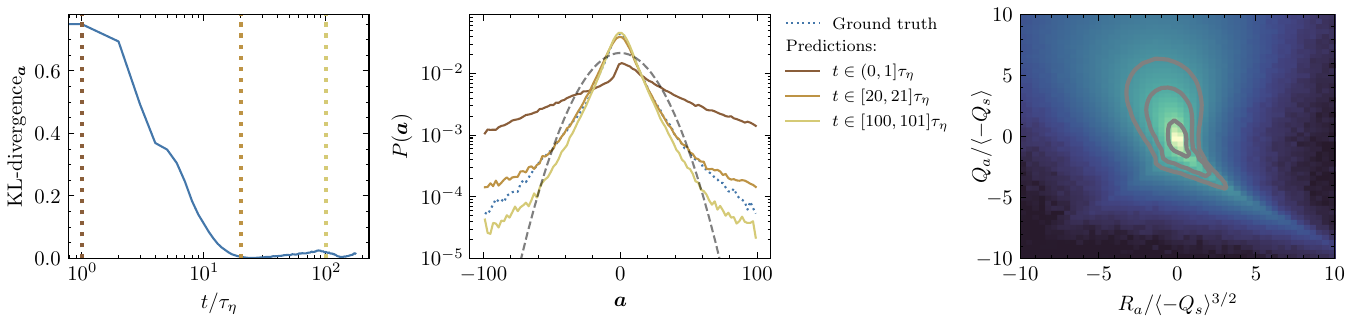}
     \put(-1,  22){\textbf{ (a) }}
     \put(28,  22){\textbf{ (b) }}
     \put(69.7,  22){\textbf{ (c) }}
    \end{overpic}
    \caption{Model predictions when starting from statistically perturbed zero initial conditions. (a) KL-divergence $D_{\textrm{KL}}(\textrm{ground-truth}\, || \,\textrm{model})$ based on the distribution of particle acceleration as a function of time throughout the predicted trajectories. (b) corresponding PDFs of acceleration for selected time windows. The dashed line indicates a Gaussian distribution with the same variance. (c) Q-R diagram of the predicted velocity gradient at the particle position, averaged from $t>20\tau_\eta$.}
    \label{fig:start_from_zero}
\end{figure*}

\subsection*{Generating trajectories from noise}

So far, all the analyses presented have assumed that the initial conditions, as well as the memory terms used to initialize the dynamics, are drawn from physically consistent data extracted from the reference dataset. We now consider a more challenging test case in which the system is initialized from a fluid at rest, with only slight perturbations introduced in the initial values of acceleration and velocity gradients to statistically separate individual particles.

In Fig.~\ref{fig:start_from_zero}a, we plot the Kullback-Leibler (KL) divergence between the model and the ground truth, evaluated at different time steps and expressed in terms of $t / t_{\eta}$. We observe that after a relatively short transient phase, lasting approximately $\mathcal{O}(10 \tau_{\eta})$, the KL divergence drops to values very close to zero. This is also reflected in the results shown in Fig.~\ref{fig:start_from_zero}b, where we plot the PDF of acceleration over three different time windows, corresponding to early-stage, intermediate, and long-time dynamics.

Remarkably, after the initial transient phase, and despite the out-of-distribution starting conditions, the model successfully drives the dynamics onto the correct attractor that reproduces the turbulent statistics. This behavior provides strong further evidence that the model has learned the correct underlying physical mechanisms governing the system dynamics.
From a practical point of view, this opens up the possibility to generate statistically adequate Lagrangian trajectories even without supplying ground truth initial conditions.

\section*{Discussion}\label{sec:conclusions}

A central challenge in modeling dynamical systems from a Lagrangian perspective lies in accurately describing the evolution of tracer particles when only a limited number of degrees of freedom are accessible, as opposed to fully resolved dynamics provided by DNS.
Such reduced representations are common in real-world scenarios; for example, when tracking pollutants transported by environmental flows, it is often only feasible to track a sparse set of particles, hence with limited information about the entire flow field. This incomplete observation complicates model fidelity, requiring careful application of embedding and closure techniques to reconstruct or infer the unobserved dynamics and enable reliable predictions.

In this work, we have demonstrated that coupling the data-driven Mori--Zwanzig formalism with time-delay embedding yields a powerful framework for learning the complex time dynamics of Lagrangian particle trajectories in turbulent flows. This technique results in a general approach with enhanced expressive capabilities, allowing the model to reconstruct the influence of unresolved degrees of freedom from sequences of past observations. In doing so, the model is able to infer the underlying phase-space structure of the dynamics without requiring explicit access to all variables of the full-order system.

The resulting reduced-order model leverages intrinsic memory of the system to accurately capture multiscale phenomena. We show that by incorporating these memory effect and training the model over many trajectories with a short-time horizon -- on the order of the Kolmogorov time scale -- it is possible to capture the correct physical mechanisms driving the dynamics.

As a result, the model achieves statistically accurate predictions over significantly longer timescales. This is demonstrated through comparisons of the PDF of acceleration and the autocorrelation of acceleration against ground truth data from DNS. We also validate the model performance by examining the correlations with local velocity gradients. Furthermore, the model exhibits robust behavior even when initialized from out-of-distribution conditions, such as zero velocity perturbed by noise, highlighting its ability to self-correct and reconstruct realistic trajectories.

We point out that, owing to the universality of small-scale turbulence, this model also holds potential to be used for turbulent flows away from the idealized setting of homogeneous isotropic turbulence. The advection due to resolved mean flows could then be superposed onto the turbulent fluctuations provided by the model in this work to yield an effective Lagrangian large-eddy simulation. In future work, we plan to evaluate the capability of the model for training at higher Reynolds number, extending to larger datasets, and further expanding the generalizability of our model.

This study opens up several promising new avenues for future applications. One particularly relevant scenario involves the real-time control of active particles in turbulent environments~\cite{calascibetta2023optimal,jiao-natcom-2025}. In such contexts, implementing model predictive control requires rapid and repeated evaluation of the system response. Running fully resolved flow simulations in real time is computationally prohibitive, whereas evaluating a reliable surrogate offers a practical and efficient alternative. By capturing the essential dynamics with significantly reduced computational cost, data-driven reduced-order models could enable real-time feedback control and decision-making in complex, unsteady flow environments.

\matmethods{

\subsection*{The Mori-Zwanzig formalism}\label{sec:mz}

For completeness, and to fix notation, we provide a summary of the Mori-Zwanzig formalism and the data-driven method used in this work. Originally introduced in statistical mechanics nearly sixty years ago~\cite{mori1965transport, zwanzig1973nonlinear}, the Mori--Zwanzig (MZ) formalism provides a rigorous mathematical framework for constructing non-Markovian reduced-order models, capturing the influence of unresolved variables through a memory kernel and orthogonal dynamics~\cite{mori1965transport, zwanzig1973nonlinear}.
The resulting reduced-order model, known as the Generalized Langevin Equation (GLE), is an integro-differential equation that governs the dynamics of selected observables (resolved variables) and includes a Markovian term, a memory-dependent term, and an orthogonal dynamics (or noise) term~\cite{Grabert1982}. This projection operator formalism, and the resulting GLE, can be found in a variety of applications, including fluid mechanics~\cite{karthik_2017, Falkena_2019} and molecular dynamics~\cite{zhen_md_2015, P_Espanol_1995}.  The memory term quantifies the interactions between resolved and unresolved dynamics. This memory effect is highly sensitive to both the choice of observables and the projection operator, and thus classical approaches~\cite{chorin2005, zhu_mori-zwanzig_2019} to analytical derivation of the Mori--Zwanzig operators have been especially challenging.

Recent advances have led to the development of data-driven Mori--Zwanzig methodologies~\cite{lin2021datadriven_full, woodward2023mzmd, zhu2021mz}, extending approximate Koopmanian learning and demonstrating improved performance compared to traditional methods such as Dynamic Mode Decomposition (DMD), Extended DMD (EDMD) and Higher-Order DMD. Furthermore, by considering regression as the projection operator~\cite{lin-siam-2023,lin-siam-2023} $\mathcal{P}$, nonlinear (neural network) based models for the MZ operators have been formalized.  These approaches have shown promising results in applications involving stationary homogeneous isotropic turbulence~\cite{Tian_2021} and boundary layer transition phenomena~\cite{woodward23_aviation, woodward23_scitech, woodward2023mzmd} as well as transient flows~\cite{woodward2024mori}. 

We now provide a general outline of the MZ framework and the data-driven method following the work of \cite{lin-siam-2023}. The goal of the Mori--Zwanzig procedure is to construct evolutionary equations for a reduced set of observables $\bm{g} \in \mathbb{R}^M$, $M\ll N$, which can be any $L^2-$integrable function of the state, also referred to as the resolved variables, where the full state $\bm p \in \mathbb{R}^N$ is intractable to resolve. The Mori--Zwanzig formalism then proceeds with a specified projection operator $\mathcal{P}$, which maps a function of the full-space configuration, $\bm{g}:\mathbb{R}^N \rightarrow \mathbb{R}^M$, to a function of only the resolved observables $\mathcal{P} \bm{g}:\mathbb{R}^M \rightarrow \mathbb{R}^M$. An operator algebraic derivation results in the generalized Langevin equation (GLE)~\cite{zwanzig1973nonlinear}:
\begin{equation}\label{eq:GLE}
    \frac{\mathrm{d}}{\mathrm{d} t} \bm g(t) = \bm M (\bm g(t)) - \int_{0}^t \bm K(t-s, \bm g(s))\mathrm{d}s + \bm F(t),
\end{equation} 
which governs the time evolution of the resolved observables $\bm g(t)$ through the Markovian term $\bm{M}$ and memory term $\bm{K}$. The GLE is a closed and exact integro-differential equation for the resolved variables, although it is now non-local in time and contains an orthogonal dynamics term $\bm F$ (often referred to as noise term) that depends on time and the initial full state. 

The GLE can then be written in discrete-time~\cite{lin_wiener_21, darve_09, GILANI2021, She_2023}, which prescribes the exact and closed set of non-Markovian evolutionary equations for the observables given any initial condition of states as:
\begin{equation}\label{eq:disc_gle}
    \bm g_{n+1} = \bm \Omega_0 (\bm g_{n}) + \sum_{k=1}^n \bm \Omega_k(\bm g_{n-k}) + \bm W_n.
\end{equation} 
The discrete time GLE \eqref{eq:disc_gle} states that the vector of observables $\bm{g}_{n+1}$ at time $n+1$ evolves according to three parts: (1) a \textit{Markovian} operator: $\bm{\Omega}_0: \mathbb{R}^M \rightarrow \mathbb{R}^M$ which only depends on the observables at the previous time step $n$, (2) the \textit{memory kernel}: the series of operators $\bm{\Omega}_k: \mathbb{R}^M \rightarrow \mathbb{R}^M$ depending on observables with a time lag $k$, and (3) the \textit{orthogonal dynamics}: $\bm W_n : \mathbb{R}^N \rightarrow \mathbb{R}^M$ depending on the full (initial) state. The above GLE is general for any projection operator (see~\cite{lin-siam-2023} for a detailed discussion).

The operators $\bm \Omega_k$ and $\bm W_n$ depend on the choice of the projection operator $\mathcal{P}$, the choice of observables $\bm g$, and the finite-time ($\Delta$) Koopman operator $\mathcal{K}$ ~\cite{li2017computing, lin_wiener_21}:
\begin{subequations}\label{eq:projection}
\begin{align}
    \bm \Omega_n &:= \mathcal{P}\mathcal{K}[(1 - \mathcal{P})\mathcal{K} ]^n, \label{eq:omega_rel} \\
    \bm W_n &:= [(1 - \mathcal{P})\mathcal{K} ]^{n+1} \bm g . \label{eq:w_rel}
\end{align}
\end{subequations}

An essential aspect of the Mori--Zwanzig formalism, often overlooked in modeling papers, is that the memory kernel and orthogonal dynamics are not independent. This general relationship between the memory kernel $\bm \Omega$ and the two-time correlation of orthogonal dynamics $\bm W$, commonly known as the generalized fluctuation-dissipation (GFD) relationship \cite{Kubo_1966, lin_wiener_21, lin2021datadriven_full} can be expressed in the discrete form as:
\begin{equation}\label{eq:gfd}
    \bm \Omega_n \bm g = \mathcal{P} \mathcal{K} \bm W_{n-1}, \quad \forall n\in \mathbb{N},
\end{equation}
where $\mathcal{K}$ is the discrete time Koopman operator. This can be derived directly from \eqref{eq:projection}.
The approximation, ultimately, then lies in neglecting the orthogonal dynamics $\bm{W}_n$ in \eqref{eq:disc_gle}, yielding the so-called truncated GLE. This ensures that the dynamics becomes a closed system that depends solely on the resolved variables $\bm{g}$ themselves, as enabled through the projection operator.

In summary, there are three key elements that one should aim to achieve with a data-driven MZ procedure: (1) learn the operators that satisfy the GLE, (2) use a well-defined projection operator, and (3) satisfy the GFD.

By extending the Dyson identity we utilize a discrete-time GLE that accommodates configurations sampled at irregular intervals of time. The Dyson identity is used in the derivation of the GLE. It is an operator identity for splitting the exponential
of a sum of operators. For two operators $A$ and $B$, it states
\begin{equation}
  e^{t(A+B)}
  = e^{tA} + \int_0^t d\tau\, e^{(t-\tau)\left(A+B\right)}\, B\, e^{\tau A}.
\end{equation}
In the context of the Mori--Zwanzig, one
decomposes the Liouville operator $\mathcal{L}$ as
\begin{equation}
  \mathcal{L} = \mathcal{QL} + \mathcal{PL},
\end{equation}
where $\mathcal{Q} = 1 - \mathcal{P}$.
The Dyson identity then takes the form
\begin{equation}
  e^{t \mathcal{L}}
  = e^{t \mathcal{QL}}
    + \int_0^t d\tau\, e^{(t-\tau)\mathcal{L}}\, \mathcal{P}  \mathcal{L}\, e^{\tau \mathcal{QL}}.
\end{equation}
In discrete form, the Dyson identity becomes~\cite{darve_09, lin_wiener_21}:
\begin{equation}
  \mathcal{K}^{n+1}
  = \sum_{k=0}^{n} \mathcal{K}^{n-k} \mathcal{PK}(\mathcal{QK})^k + (\mathcal{QK})^{n+1},
\end{equation}
where $\mathcal{K}$ is the discrete time Koopman operator $\mathcal{K} \equiv e^{\Delta t \mathcal{L}}$.
Specifically, the projection operators associated with each sampling instance can differ, unlike conventional MZ treatments that assume a single fixed projector. This generalization enables a data-driven construction with variable time steps.

Several data-driven methods have been proposed for estimating the MZ operators ~\cite{chorin_optimal_2000, chorin_optimal_2002, lin_wiener_21, li2017computing, gonzalez_learning_2020, zhu_estimation_2018, lin-siam-2023}. In the data-driven MZ method \cite{lin-siam-2023} used in this work, we assume a finite length memory $h$. The MZ operators $\bm \Omega_k$ (for $k \in \{0, 1, ..., h\}$) in the truncated GLE are learned with Neural Networks, enforcing the GFD relation by following Ref.~\cite{lin-siam-2023}, and using regression as the projection operator. Additionally, we take advantage of the variable projection method in order to explore a GLE with variable time steps. For more details on the derivation and algorithms, the reader is referred to Ref.~\cite{lin-siam-2023}.

\subsection*{Time-delay embedding}\label{sec:takens}

To further increase the expressiveness of the model, the MZ operators can be extended by incorporating additional time-delay embeddings. In this formulation, every memory kernel acts not only on the resolved state at a given time but also on an augmented state vector that includes a finite sequence of past states, referred to as time-delay embeddings. The theoretical foundation for this approach was rigorously formalized by Takens’ embedding theorem \cite{takens81}.

Since MZ memory provides a general term to account for how the unresolved variables interact with the resolved observables, MZ memory and delay embeddings are not mutually exclusive. Integrating time-delay embedding with the Mori–Zwanzig formalism offers a natural and systematic strategy to address unresolved scales or variables, explicitly introducing memory effects while simultaneously enlarging the functional space for regression. For example, Lin et al.~\cite{lin-siam-2023} showed that their combination can produce optimal predictive performance.
Including MZ memory can thus alleviate the need to determine the embedding dimension of the strange attractor assumed by Takens' theorem. Consequently, this combined framework can produce more accurate, robust, and interpretable reduced-order models.

\subsection*{Data-driven Mori--Zwanzig for Lagrangian turbulence}\label{sec:ml}

We now introduce the data-driven Mori--Zwanzig formalism augmented with time-delay embedding in the context of Lagrangian turbulence. The resulting reduced-order model explicitly governs the temporal evolution of particle acceleration $\bm{a}$; where particle position and velocity are directly obtained via temporal integration.  
Furthermore, we include local fluid velocity gradients $\bm {\nabla}\bm{u}$ at the position of the particle in the model to encode the knowledge of local small-scale structures of turbulent flow~\cite{jiao-natcom-2025}, such as strain and vorticity, which strongly influence particle dynamics. Formally, the model is then defined as
\begin{subequations}\label{eq:mz-basic-all}
\begin{align}
    \begin{pmatrix} 
        \bm{a}_{n+1} \\ 
        \bm{\nabla}\bm{u}_{n+1}
    \end{pmatrix} 
    &= 
    \textrm{MZ}_h(\bm{a}_{n,...,n-h}, \bm {\nabla}\bm{u}_{n,...,n-h}),\label{eq:mz-basic}\\
    \bm{v}_{n+1} &= \bm{v}_n + \bm{a}_n \Delta t,\\
    \bm{x}_{n+1} &= \bm{x}_n + \bm{v}_n \Delta t,
\end{align}
\end{subequations} 
where the right-hand side of \eqref{eq:mz-basic} consists of the superposition between a Markovian kernel and a finite number $h$ of memory kernels as
\begin{eqnarray}
    \textrm{MZ}_h(\bm{a}_{n,...,n-h}, \bm{\nabla}\bm{u}_{n,...,n-h}) 
    = 
    \overbrace{\bm \Omega_0(\bm{a}_n, \bm{\nabla}\bm
     {u}_n)}^{\textrm{Markovian kernel}} \nonumber\\ 
    + \sum_{k=1}^h \underbrace{ \bm \Omega_k(\bm{a}_{n-k}, \bm{\nabla}\bm{u}_{n-k})}_{\textrm{memory kernels}}.
\end{eqnarray} 

The next key ingredient is the data-driven learning of the MZ operators $\bm{\Omega}_k$. Following the approach described in Ref.~\cite{lin-siam-2023}, we define the projection operator in the context of regression. Specifically, the method consists of incrementally minimizing the orthogonal dynamics (i.e., residual noise) by explicitly enforcing the generalized fluctuation–dissipation relation, \eqref{eq:gfd}, a necessary condition for consistency with the MZ formalism.
By interpreting regression as the process of forming the projection operator, and leveraging the property that the projection $\mathcal{P}$ annihilates the noise term ($\mathcal{P} \bm{W} = 0$), we can express the learned model directly in terms of projected quantities, without explicitly modeling the noise. However, in this case, it is not exactly closed, but only approximately so by the memory kernels. 

In practice, we perform the regression through deep learning, where we learn the kernels $\bm{\Omega}_k$, parameterized as fully connected multi-layer perceptron (MLP) neural network (NN). Before going in details on the training procedure, we consider a generalization of \eqref{eq:mz-basic} where the MZ paradigm is coupled with $d$ Takens' time-delay embeddings. The resulting formulation is then:
\begin{eqnarray}
    &&\textrm{MZ}_h(\bm{a}_{n,...,n-\kappa(h)-d}, \bm{\nabla}\bm{u}_{n,...,n-\kappa(h)-d}) 
    = \nonumber\\
    &&\quad\Omega_0(\bm{a}_{n,...,n-d}, \bm{\nabla}\bm{u}_{n,...,n-d}) \\ 
    &&+ \sum_{k=1}^h \Omega_k(\bm{a}_{n-\kappa(k),...,n-\kappa(k)-d}, \bm{\nabla}\bm{u}_{n-\kappa(k),...,n-\kappa(k)-d}).\nonumber
\end{eqnarray}
In the above we also allow for non-linearly time-spaced memory kernels, via the mapping $\kappa(k)$.

In the DNS, the mean flow is typically explicitly enforced to be zero, which ensures that both Eulerian and Lagrangian velocities always remain statistically bound with some spread around a zero mean. However, this zero mean condition is not inherently enforced in the surrogate model, leading to the possibility of unbounded, diffusively growing velocities. To enforce a zero mean condition, we thus apply a small linear damping based on the Lagrangian velocity to the predicted acceleration, replacing \eqref{eq:mz-basic} by
\begin{equation}\label{eq:mz-damped}
    \begin{pmatrix} 
        \bm{a}_{n+1} \\ 
        \bm{\nabla}\bm{u}_{n+1}
    \end{pmatrix} 
    = \textrm{MZ}_h(\bm{a}_{n,...,n-h}, \bm{\nabla}\bm{u}_{n,...,n-h}) 
    - 
    \begin{pmatrix} 
        \alpha \bm{v}_n \\ 
        \bm{0} 
    \end{pmatrix}.
\end{equation}
The damping factor $\alpha$ should be tuned in such a way that its action remains restricted to the longest time scales. In practice this means that the damping time scale $\alpha^{-1}$ should be on the order of the integral timescale of turbulence to ensure that the short time and inertial range dynamics remain unaffected. The numerical analysis of the effect of the damping term is provided in SI Appendix~B.

\subsection*{Memory in Lagrangian turbulence}
Temporal correlations in Lagrangian trajectories in turbulence can in principle extend from the \mbox{(sub-)Kolmogorov} scales all the way to the integral time scales. This offers a practical challenge for modeling high-Reynolds flows, as one needs to take correlations into account over a large dynamic range of scales. We therefore opt for a logarithmic spacing of memory terms $\kappa(k)=\lambda^{k-1}$, where we choose $\lambda=2$, allowing us to account for correlation over a large range of time-scales whilst limiting the number of memory terms. Since the accelerations that we are modeling are related to the velocity gradient field, they are dominated by contributions from the small scales. Hence, to maximize the information from the short-time correlations, we keep the Takens' time-delay embedding linearly spaced, maintaining a high resolution of the short-time history, on the order of the Kolmogorov scale.

\subsection*{Hyper-parameters and Loss Function}

In order to define the training procedure for establishing the kernels $\bm{\Omega}_k$ discussed in the previous section, we introduce the loss function, defined as the sum of the mean-squared error computed on the prediction of particle acceleration and fluid velocity gradient:
\begin{equation}
   \mathcal{L} = \mathrm{MSE}_{\bm{a}} + \mathrm{MSE}_{\bm{\nabla} \bm{u}} .
\end{equation}

We employ here a technique known in the literature as \textit{ unrolled training} or \textit{trajectory learning}~\cite{brandstetter-arxiv-2022}, which consists of applying the model to make predictions up to $L_T$ consecutive time steps at training time, backpropagating the gradients through all (or just a few of) such time steps. 
We weight the contributions of each of these time steps in the loss, giving more importance to shorter-term predictions, following the same approach as put forward in Ref.~\cite{xue2024card}.
We thus train the model on the following error metric, with $\mathrm{MSE}$ defined for a generic observable $\bm{g}$ as
\begin{equation}\label{eq:nn-loss}
  \mathrm{MSE}_g = \sum_{i=1}^{L_T} w_i \left( \bm{\tilde{g}}_i - \bm{g}_i \right)^2 ,
\end{equation}
where $\bm{\tilde{g}}_i$ is the prediction of the model, and weights $w_i$ decay exponentially with rate factor $\gamma$ as \mbox{$w_i = \exp(-\gamma i / L_t)$}. The unroll horizon $L_t$ should be chosen on the order of the Kolmogorov time, such that one can expect to be able to make predictions that are point-wise accurate on this time scale, minimizing the mean-squared error.
Optimization is done with stochastic gradient descent using the Adam optimizer and with early stopping based on validation loss to avoid overfitting. 

The full choice of all hyper-parameters employed in this work is provided in Tab.~\ref{tab:hyper}. 
Moreover, in SI Appendix~A, we report an analysis of the impact of memory lenghts and unroll horizon parameters.

\begin{table}[h!]\centering
    \vspace{6mm}
    \caption{\label{tab:hyper}
    Hyper-parameters used for the MZ model.}
    \begin{tabularx}{0.9\linewidth}{Xr}
        \hline\\
            Number of hidden layers & 9 \\
            Neurons per layer & 768 \\
            Model size & 5M \\
            Total number of weights & $5\times 10^6$ \\
            Number of epochs & 500 \\
            Unroll horizon $L_T$ & 5 \\
            MZ history $h$ & 5 \\
            Delay embeddings $d$ & 40 \\
            Time-step $\Delta t$ & $0.1\tau_\eta$ \\
            Damping $\alpha$ & 0.5 \\\\
        \hline\hline
    \end{tabularx}
\end{table}

}

\showmatmethods{} 

\acknow{This work was supported in part by the Netherlands Organization for Scientific Research (NWO) through the use of supercomputer facilities (Snellius) under Grant No. 2023.026, 2025.008. This publication is part of the project “Shaping turbulence with smart particles” with Project No. OCENW.GROOT.2019.031 of the research program Open Competitie ENW XL which is (partly) financed by the Dutch Research Council (NWO).
This work is co-authored by employees of Triad National Security, LLC which operates Los Alamos National Laboratory (LANL) under Contract No.89233218CNA000001 with the U.S. Department of Energy/National Nuclear Security Administration. A.G. gratefully acknowledges the support of the LANL/LDRD Program under project number 20240740PRD and the Center for Non-Linear Studies for this work.
Y.T.L.~and D.L.~were supported by the U.S.~Department of Energy, Office of Science, Office of Advanced Scientific Computing Research’s Applied Mathematics Competitive Portfolios program. X.W., M.W., and D.L. also acknowledge the support from the LANL/LDRD Program under project number 20240318ER. }

\showacknow{} 


\bibliography{main}

\end{document}



\maketitle

\SItext

\section*{Appendix A: Ablation study}\label{app:ablation}

\begin{figure}
    \centering
    \includegraphics[width=.45\linewidth]{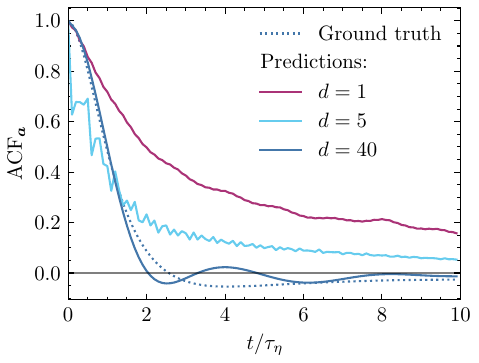}
    \caption{Influence of number of time-delay embeddings $d$, depicted through the autocorrelation of acceleration (with $h=5$ memory kernels). More delay embeddings yield a better prediction.}
    \label{fig:params_delay}
\end{figure}

In this Appendix section, we systematically investigate the influence of the most important hyperparameters on the statistical accuracy of long-time predictions produced by the MZ model. Specifically, we focus on three aspects: (i) the number of time-delay embeddings $ d $ (ii) the number of memory kernels $ h $ and (iii) the length of the unroll horizon $ L_T $ used during training.

\begin{figure}
    \centering
    \includegraphics[width=.45\linewidth]{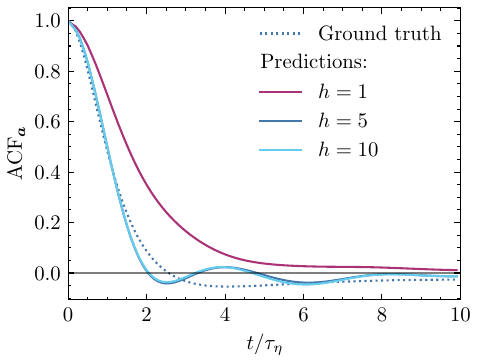}
    \caption{Influence of number of memory kernels $h$, depicted through the autocorrelation of acceleration (with $d=40$ time-delay embeddings). We find no significant improvement beyond $h\approx 5$.}
    \label{fig:params_mzhistory}
\end{figure}

All results presented here are obtained using a reduced training dataset consisting of 500K Lagrangian trajectories, and the model's performance is evaluated based on its ability to reproduce the autocorrelation and PDF of particle accelerations over extended time horizons.

First, we discuss the effect of time-delay embeddings. We find that time-delay embeddings are particularly important for properly resolving the time-correlation structure of the Lagrangian trajectories. In Fig.~\ref{fig:params_delay}, we examine how varying the number of embeddings $ d $ affects the temporal autocorrelation of the predicted particle acceleration. We observe that increasing $ d $ leads to progressively better agreement with the ground truth, with the autocorrelation function approaching the DNS reference as $ d $ increases. This result confirms the effectiveness of delay embedding in resolving the intrinsic memory of the system. However, the added input dimensionality also increases model complexity and computational cost. We find that using approximately $ d \approx 40 $ delay steps, corresponding to a temporal window of $ 4\tau_\eta $ given the sampling interval $ \mathrm{d}t \approx 0.1 \tau_\eta $, provides a good trade-off between accuracy and efficiency.

Next, we assess the impact of the number of memory kernels $h$ which control how many past time instances the model explicitly retains through the MZ formalism. In this test we vary $h$ keeping $d$ fixed to $d=40$. As for the time-delay embeddings, we find that the memory kernels are particularly important for properly resolving the time-correlation structure.
As shown in Fig.~\ref{fig:params_mzhistory}, we see that increasing $h$ improves the accuracy of acceleration autocorrelations. However, we find that beyond $h\approx 5$ additional memory kernels 
do not significantly improve the results, suggesting that the memory architecture can remain relatively shallow without sacrificing predictive accuracy, provided the embeddings are sufficiently rich.

\begin{figure}
    \centering
    \begin{overpic}[width=.45\linewidth]{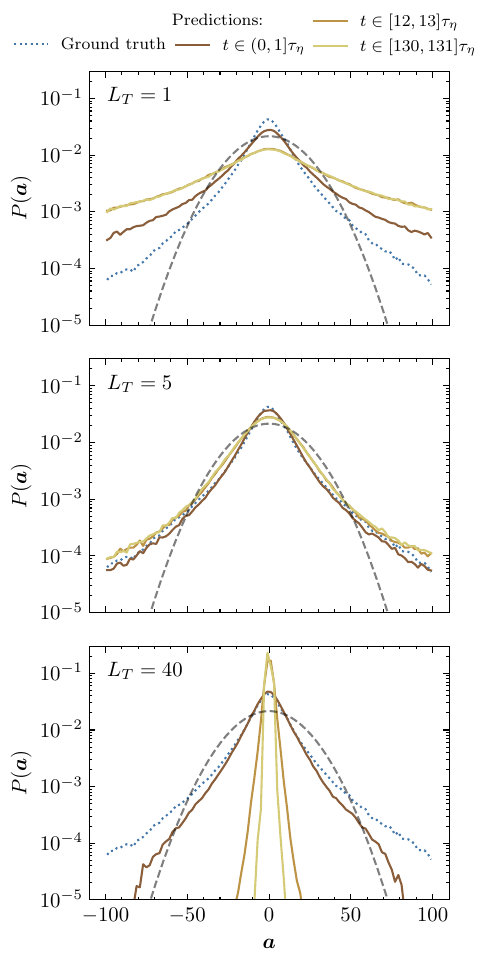}
     \put(0,  92){\textbf{ (a) }}
     \put(0,  62){\textbf{ (b) }}
     \put(0,  32){\textbf{ (c) }}
    \end{overpic}
    \caption{Influence of the unroll horizon in training $L_T$, depicted through PDFs of acceleration at different time intervals throughout the predicted trajectory for different number of unroll steps $L_T=1$ (a), $L_T=5$ (b) and $l_T=40$ (c). The dashed lines indicate Gaussians with the same variance.}
    \label{fig:params_unroll}
\end{figure}

Finally, we discuss the influence of the unroll horizon $L_T$ used during training, which determines how many future steps are included in the training loss function.
We find that unrolled training is particularly influencing the robustness of the long term prediction. In Fig.~\ref{fig:params_unroll}, we show the predicted distribution of acceleration at different time intervals throughout the prediction for different values of $L_T$. We find that there is an optimum number of unroll steps around $L_T \approx 5$ (panel b). Having a too short unroll horizon (panel a) results in increasingly extreme predictions of acceleration, while having too many unroll steps (panel c) results in increasingly conservative predictions of the acceleration around the mean value of zero.

\FloatBarrier

\section*{Appendix B: Velocity damping} \label{app:damping}

\begin{figure}
    \centering
    \begin{overpic}[width=.45\linewidth]{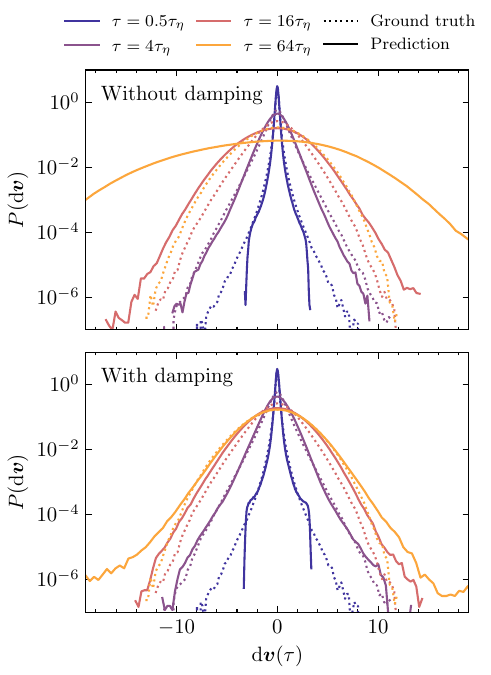}
     \put(0,  91){\textbf{ (a) }}
     \put(0,  48){\textbf{ (b) }}
    \end{overpic}
    \caption{PDFs of Lagrangian velocity differences $\mathrm{d}\bm{v}(\tau)$ for various time lags $\tau$, comparing the model without damping $\alpha=0$ (a) and with damping $\alpha=0.5$ (b).}
    \label{fig:damping}
\end{figure}

In Fig.~\ref{fig:damping}, we investigate the effect of including the damping term in the model, as for \eqref{eq:mz-damped}. For this analysis, we take into consideration the PDF of the velocity increment $\mathrm{d} \bm{v} (\tau) = \bm{v}(t + \tau) - \bm{v}(t)$.

The figure compares the results at different time lags, specifically $\tau = 0.5 \tau_{\eta}, 4 \tau_{\eta}, 16 \tau_{\eta}$, and $64 \tau_{\eta}$. The top panel shows the results for the model without damping (i.e., $\alpha = 0$, see \eqref{eq:mz-damped}), while the bottom panel displays the results obtained by training the model with $\alpha = 0.5$.

Without damping, we observe that, indeed, the particle velocities tend to grow without bound at larger time lags, as is expected from contributions coming from unbounded diffusion, as explained above.
In contrast, the inclusion of the damping term effectively mitigates the problem, and the PDFs remain much closer to those of the ground truth across all time scales considered, thereby mimicking the zero mean flow condition that is enforced in the DNS.

\section*{Appendix C: Velocity gradient PDFs and auto-correlations}\label{app:vg}

\begin{figure}
    \centering
    \begin{overpic}[width=.9\linewidth]{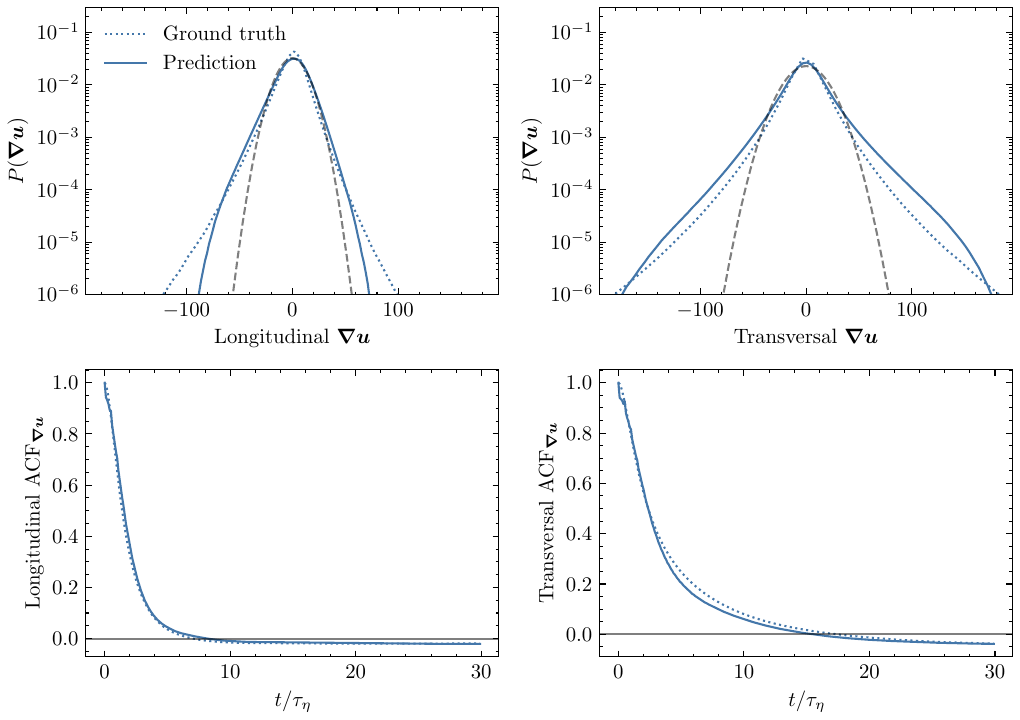}
     \put(0,  69){\textbf{ (a) }}
     \put(50,  69){\textbf{ (b) }}
     \put(0,  34){\textbf{ (c) }}
     \put(50,  34){\textbf{ (d) }}
    \end{overpic}
    \caption{Statistics of the long time prediction by the MZ model of the velocity gradients. Provided are the PDFs (a,b) and autocorrelation functions (c,d) of longitudinal (a,c) and transversal (b,d) velocity gradients. The dashed lines in de PDFs indicate Gaussians with the same variance.}
    \label{fig:vg_stats}
\end{figure}
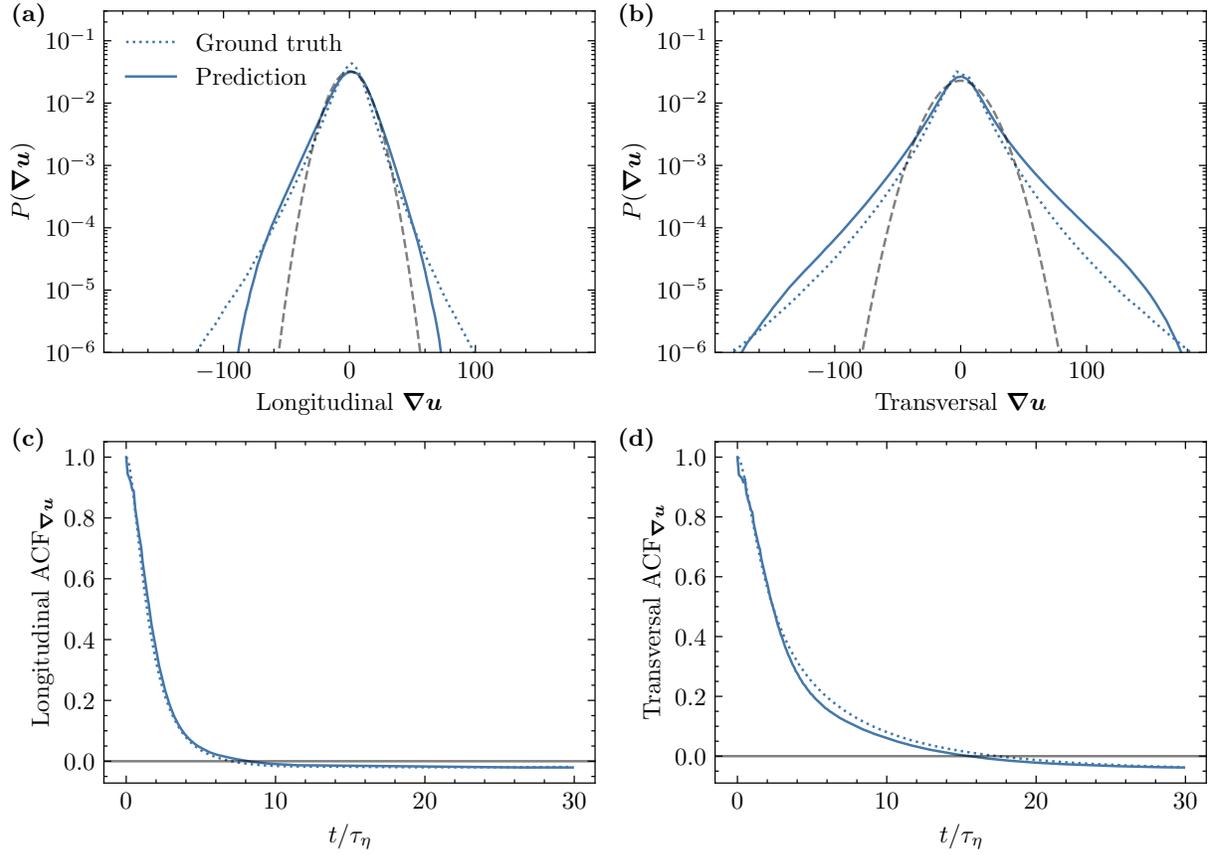

In this Appendix section we complement the analysis presented in the main text (cf. Fig.~\ref{fig:best_stats}) examining the statistical properties of the velocity gradient tensor predicted by the MZ model over long time horizons. Fig.~\ref{fig:vg_stats} presents the PDFs and autocorrelation functions of both longitudinal and transverse components of the velocity gradient, computed from long-term model-generated trajectories.
In the top row, the PDFs of the longitudinal and transverse velocity gradients are shown, highlighting the capability of the model to describe the strongly non-Gaussian nature of these statistics.

The bottom row shows the corresponding temporal autocorrelation functions, which provide insight into the memory and coherence timescales of the gradient components. The decay rates predicted by the model are in excellent agreement with those from DNS data, indicating that the learned dynamics preserve the correct temporal structure of small-scale velocity fluctuations.

\section*{Appendix D: Comparison to model without memory}\label{app:nomem}
\begin{figure}
    \centering
    \begin{overpic}[width=.9\linewidth]{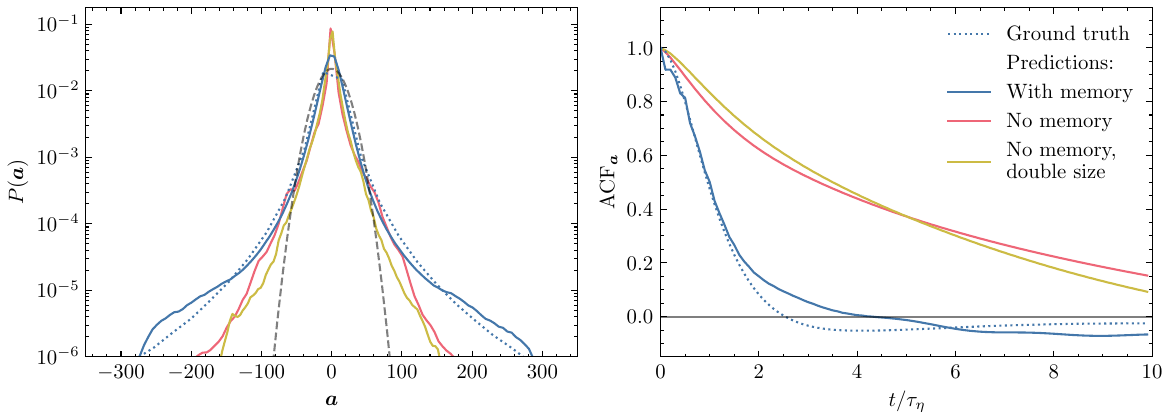}
     \put(0,  36){\textbf{ (a) }}
     \put(50,  36){\textbf{ (b) }}
    \end{overpic}
    \caption{Comparison between the performance of the memory-based approach proposed in the main text and a memoryless approach. We compare the PDF of acceleration (a) and the corresponding auto-correlation (b) for the memory-based model treated in the main text (blue line), a memoryless approach with the same network size (5M parameters per kernel, red line) and a memoryless approach with double the network size (10M parameters per kernel, yellow line).}
    \label{fig:best_stats_nomem}
\end{figure}

In this appendix, we further explore the comparison between the proposed approach that incorporates memory based on the Mori-Zwanzig formalism and Takens' time-delay embeddings and a fully memoryless approach. In the main text, we compare the proposed approach to a memoryless approach where the neural networks have the same size and the same number of kernels are used. In the memoryless case, all kernels purely Markovian, taking as input only the last time-step and without enhancing the state with time-delay embeddings. As shown in the main text, this comparison indicates that the memoryless approach in this baseline comparison is not able to capture the statistics of the turbulent dynamics, particularly the time-correlation structure. Here, we take this one step further, showing that even when considering yet a larger network size for the memoryless model, doubling the size of the neural networks of each kernel, the memoryless model is not able to improve satisfactorily towards the performance obtained by the (smaller in size) memory-based approach that we propose in the main text. These results are provided in Fig.~\ref{fig:best_stats_nomem}.

\FloatBarrier
